\documentclass[11pt]{article}
\usepackage[top=1in, bottom=1in, left=1in, right=1in]{geometry}
\usepackage{graphics,graphicx}
\usepackage{caption}
\usepackage{amssymb,amsmath,amstext,mathtools}
\usepackage{color}
\usepackage{times}
\usepackage{cite}
\usepackage{setspace}
\newcommand{\lyxaddress}[1]{
\par {\raggedright #1
\vspace{1.4em}
\noindent\par}
}

\usepackage[titletoc,title]{appendix}
\usepackage{titlesec}
\titleformat{\chapter}[display]% NEW
    {\large\bfseries}{\chaptertitlename\ \thechapter}{5pt}{\Large}% NEW
\titlespacing*{\chapter}{5pt}{5pt}{5pt}

\begin{document}

\title{Cell-cycle coupled expression minimizes random fluctuations in gene product levels}

\author{Mohammad Soltani$^{\text{}1}$ and Abhyudai Singh$^{\text{}1,2,3,*}$}

\date{}

\maketitle

\lyxaddress{1. Department of Electrical and Computer Engineering, University of Delaware, Newark, DE, USA.\\
2. Department of Mathematical Sciences, University of Delaware, Newark, DE, USA.\\
3. Department of Biomedical Engineering, University of Delaware, Newark, DE, USA. \\
}

\lyxaddress{{*} Corresponding Author: Abhyudai Singh,  \texttt{absingh@udel.edu}.}

\newpage
\section*{Abstract}
Expression of many genes varies as a cell transitions through different cell-cycle stages. How coupling between stochastic expression and cell cycle impacts cell-to-cell variability (noise) in the level of protein is not well understood. We analyze a model, where a stable protein is synthesized in random bursts, and the frequency with which bursts occur varies within the cell cycle. Formulas quantifying the extent of fluctuations in the protein copy number are derived and decomposed into components arising from the cell cycle and stochastic processes. The latter stochastic component represents contributions from bursty expression and errors incurred during partitioning of molecules between daughter cells.  These formulas reveal an interesting trade-off: cell-cycle dependencies that amplify the noise contribution from bursty expression also attenuate the contribution from partitioning errors. We  investigate existence of optimum strategies for coupling expression to the cell cycle that minimize the stochastic component. Intriguingly, results show that a zero production rate throughout the cell cycle, with expression only occurring just before cell division minimizes noise from bursty expression for a fixed mean protein level.  In contrast, the optimal strategy in the case of partitioning errors is to make the protein just after cell division. We provide examples of regulatory proteins that are expressed only towards the end of cell cycle, and argue that such strategies enhance robustness of cell-cycle decisions to the intrinsic stochasticity of gene expression.

\newpage
\section{Introduction}

Advances in experimental technologies over the last decade have provided important insights into gene expression at a single-molecule and single-cell resolution. An important (but not surprising) revelation is the stochastic expression of genes inside individual cells across different organisms \cite{bkc03,rao05,keb05,nmt_13,mal13,jbp14,sgz11,smg11,drs12,singh_transient_2014,hbr12}. In many cases, stochastic expression is characterized by random burst-like synthesis of gene products during transcription and translation.  At the transcriptional level, promoters randomly switch to an active state, producing a burst of RNAs before becoming inactive \cite{rpt06,srd12,btl14,bhh16,dsp15,brb14}. At the translational level, a relatively unstable mRNA degrades after synthesizing a burst of protein molecules \cite{cfx06,shs08,sih09b,otk02}. Bursty expression drives intercellular variability (noise) in gene product levels across isogenic cells, significantly impacting biological pathways and phenotypes \cite{ccf08,src10,Eldar:2010kk,siw09,sid13,rao08,ulo16,nlp15}.

Mathematical models have played a key role in predicting the impact of bursty expression on noise in the level of a given protein. However, these studies have 
primarily relied on models where synthesis rates are assumed to be constant and invariant of cell-cycle processes. While such an assumption is clearly violated for cell-cycle regulated genes, replication-associate changes in gene dosage can alter expression parameters genome wide \cite{yxr06,zopf13,sxn16,pnb15}. It is not clear how such cell-cycle dependent expression affects the stochastic dynamics of protein levels in single cells.  To systematically investigate this question, we formulate a model where a cell passes through multiple cell-cycle stages from birth to division. Cell cycle is coupled to bursty expression of a stable protein and the rate at which bursts occur depend arbitrarily on the cell-cycle stage (Fig. 1). In addition to stochastic expression in bursts, the model incorporates other physiological noise sources, such as, variability in the duration of cell-cycle times and random partitioning of molecules between daughter cells at the time of division \cite{hup11,lambert_2015,reshes_timing_2008,Roeder:2010vb,Zilman:2010ud,Hawkins:2009vw,sak13,gon13,ltr14}. 

In the proposed model, some cell-to-cell variability or noise in the protein level is simply a result of cells being in different cell-cycle stages (i.e., asynchronous population). We illustrate a novel approach that takes into account such cell-cycle effects, and quantifies the noise contribution just from bursty expression and partitioning errors. Formulas obtained using this approach reveal that cell-cycle dependent expression considerably alters noise levels, always affecting contributions from bursty expression and partitioning errors in opposite ways. Intriguingly, our results show existence of optimal strategies to synthesize a protein within the cell cycle that minimize noise contributions for a fixed mean protein level. For example, the noise contribution from bursty expression is minimal when the protein is synthesized only towards the end of cell cycle. We discuss intuitive reasoning behind these optimal strategies, and provide examples of proteins that are expressed in this fashion to enhance fidelity of cell-cycle decisions.

\begin{figure}[!ht]
\centering
\includegraphics[width=0.7\linewidth]{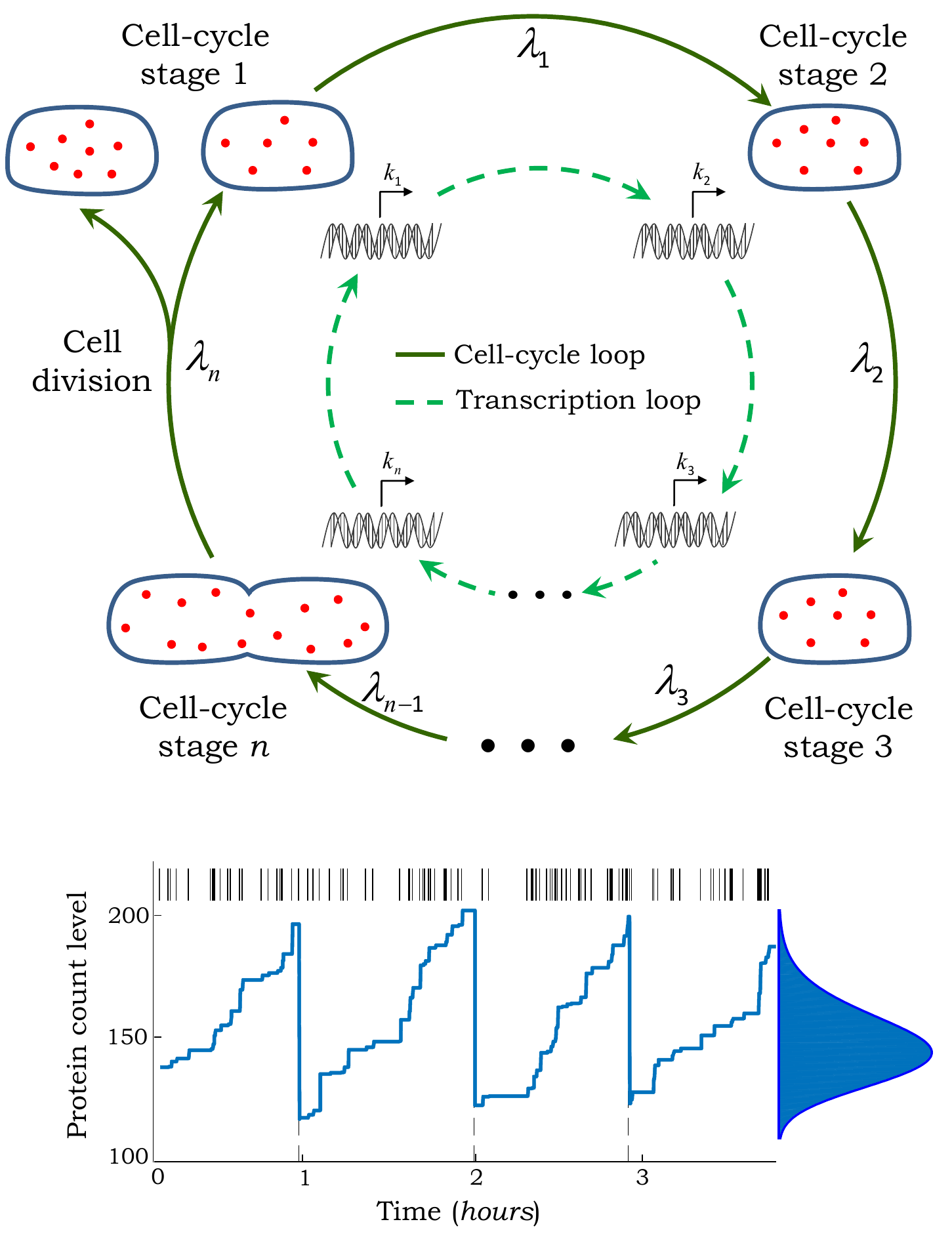}
\caption{{\bf Coupling cell cycle to gene expression}. {\it Top}: The outer loop shows an individual cell from birth to division passing through cell-cycle stages $C_1$, $C_2$, \dots, $C_n$, with transition rates between stages give by $\lambda_i$, $i\in\{1,2,\ldots,n\}$. The cell is born in stage $C_1$ and division is initiated in $C_n$. The inner loop (transcriptional cycle) represents the rate at which protein expression bursts occur and is given by $k_i$ in cell-cycle stage $C_i$. {\it Bottom}: Representative trajectory of the protein level in an individual cell through multiple cell cycles (dashed lines). In this case, the transcription rate is assumed to double at the cell-cycle midpoint due to replication-associated increase in gene dosage. The spike train above represents the firing times of burst events. 
Steady-state distribution of the protein copy numbers obtained from running a large number of Monte Carlo simulations is shown on the right. The cell cycle was modeled by choosing $n=20$ stages with equal transition rates between. Protein expression was assumed to occur in geometric bursts with 
$\langle B \rangle=10$ and molecules were partitioned between daughter cells based on a binomial distribution.}
\label{fig1}
\end{figure}

\section{Model coupling cell cycle to gene expression}

We adopt a phenomenological approach to model cell cycle and divided it into $n$ stages $C_1$, $C_2$, $\ldots$, $C_n$.  A newborn cell is in stage $C_1$ and transitions from $C_i$ to $C_{i+1}$ with rate $\lambda_i$. In stage $C_n$, cell division is initiated with rate $\lambda_n$, and upon division the cell returns to $C_1$. In the stochastic formulation of this model, the cell resides in stage $C_i$ for an exponentially distributed time interval with mean $1/\lambda_i$, and cell-cycle duration is a sum of $n$ independent, but not necessarily identical, exponential random variables. These stages can be mathematically characterized by Bernoulli processes  $c_1(t)$, $c_2(t)$, $\ldots$, $c_n(t)$, where $c_i(t)=1$ when the cell is in stage $C_i$ and $c_i(t)=0$ otherwise. Based on the model structure, these processes satisfy
\begin{align}
&\sum_{i=1}^n c_i(t)=1, \ \  c_i(t) c_j(t)=0 \ \  {\rm for} \ \ i \neq j, \label{c}
\end{align}
The latter equality results from the fact that only one of the $c_i$ can be equal to $1$ at any given time. In addition, since $c_i$ takes values in $\{0,1\}$
\begin{align}
\langle c_i^m \rangle=\langle c_i \rangle, \ \ m\in\{1,2,\ldots\}, 
\end{align}
where the symbol $\langle \ \rangle$ denotes the expected value. Next, we describe the coupling between the cell cycle and stochastic expression models. 

We assume that gene-expression bursts occur at a Poisson rate $k_i$ in cell-cycle stage $C_i$. Using the above-defined Bernoulli processes, the burst arrival rate can be compactly written as $\sum_{i=1}^n k_i  c_i(t)$. Let $x(t)$ denote the number of protein molecules in a singe cell at time $t$. Then, whenever burst events occur, the protein level is reset as 
\begin{equation}
\begin{aligned}
 x(t) \mapsto x(t)+ B ,
\end{aligned}
\label{model0}
\end{equation}
where the burst size $B \in \{0,1,2,\dots\}$ is a random variable independently drawn from an arbitrary distribution, and reflects the net contribution of transcriptional and translational bursting.  As is true for most proteins in \emph{E. coli} and \emph{S. cerevisiae}, we assume a stable protein without any active degradation between burst events \cite{alo11,tcl10,sbl11}. At the time of cell division (as dictated by the cell-cycle model), the protein molecules are randomly partitioned between daughters. This corresponds to the following reset that is activated during division
\begin{equation}
\begin{aligned}
x(t) \mapsto x_+(t),
\end{aligned}
\label{model}
\end{equation}
where the mean and variance of $x_+$ (level just after division) conditioned on $x$ (level just before division) are given by 
\begin{equation}
  \langle x_+ \vert x \rangle= \frac{x  }{2},  \ \ \ \  \left \langle{x_+^2}  \vert x\right \rangle  -\langle x_+ \rangle^2=  \frac{\alpha x}{4},
  \label{division character}
\end{equation}
respectively. The first equation in \eqref{division character} shows that the number of molecules is approximately halved during division, while the second equation quantifies the stochasticity in the partitioning process
through the parameter $\alpha$. The ideal case of zero partitioning errors corresponds at $\alpha=0$, where $x_+(t)={x (t)}/{2}$ with probability one.
Binomial partitioning, where each molecule has an equal chance of ending up in one of the two daughter cells, is given by $\alpha=1$ \cite{gpz05,berg_1978,rig79}. Finally, values of $\alpha>1$ represent additional noise in the partitioning process that arise when protein molecules form multimers, or reside in organelles that are themselves subject to binomial partitioning \cite{huh11,hup11}.  The overall model coupling cell cycle to expression is illustrated in Fig. 1 together with a representative trajectory of $x(t)$.

\section{Mean protein level for cell-cycle driven expression}

We illustrate an approach based on closing moment dynamics for deriving an exact analytical formula for the mean protein level.  The first step is to obtain differential equations describing the time evolution of the statistical moments for $x(t)$ and $c_i(t)$. These equations can be derived using the Chemical Master Equation (CME) corresponding to the stochastic model presented in the previous section (see Appendix A in SI). In particular, time evolution of the means (first-order moments) is given by 
\begin{subequations}
\begin{align}
&\frac{d\langle c_1 \rangle}{dt}=\lambda_n \langle c_n \rangle-\lambda_1 \langle c_1 \rangle, \ \ \frac{d\langle c_i \rangle}{dt}=\lambda_{c-i} \langle c_{i-1} \rangle-\lambda_i \langle c_i\rangle, \ \ i\in\{2,3,\ldots, n\}, \label{x1}\\
&\frac{d\langle x \rangle}{dt}= \left(\sum_{i=1}^n k_i \langle c_i \rangle\right)\langle B \rangle - \frac{\lambda_n}{2} \langle x c_{n} \rangle. \label{x}
\end{align} \label{x3}
\end{subequations}
Steady-state analysis of \eqref{x1} yields the average value of Bernoulli processes as
\begin{equation}
\overline{\langle c_{i} \rangle} = \frac{\frac{1}{\lambda_i}}{\sum_{j=1}^n\frac{1}{\lambda_j}}, \label{gi}
\end{equation}
which can be interpreted as the fraction of time spent in the cell-cycle stage $C_i$. We use $\overline{\langle  \ \rangle}$ to denote the expected value of a stochastic process as $t \to \infty$. 

Note that the dynamics of $\langle x \rangle$ in \eqref{x} is ``not closed", in the sense that it depends on second-order moments $\langle xc_{n} \rangle$. This leads to the well-known problem of moment closure that often arises in stochastic chemical kinetics \cite{gov06,lkk09,gou07,sih10,gil09,svs15,jdd14,sih10a}. It turns out that in this case, the model structure can be exploited to automatically close moment equations. This is done by augmenting the system of equations in \eqref{x3} with the time evolution of moments of the form $\langle xc_{i} \rangle$ 
\begin{subequations}
\begin{align}
&  \frac{d\langle x c_{1} \rangle}{dt}=k_1 \langle B \rangle  \langle c_{1} \rangle + \frac{\lambda_n}{2} \left \langle xc_{n} \right\rangle   - \frac{\lambda_n}{2}   \left \langle  {x}c_{1} c_{n} \right\rangle  - \lambda_1 \langle xc_{1} \rangle \label{gi11} ,\\
&\frac{d\langle x c_{i} \rangle}{dt}=  k_i \langle B \rangle  \langle c_{i} \rangle 
- \lambda_{i} \langle x c_{i} \rangle +\lambda_{i-1}\langle x c_{i-1} \rangle. \ \ j\in \lbrace 2,\ldots,n \rbrace. \label{gij tot}
\end{align} 
\label{gij}
\end{subequations}
At the first look, these equations are unclosed and depend on third-order moments of the form $\langle xc_i c_n\rangle $. However, exploiting the fact that $c_ic_j=0$ from 
\eqref{c} leads to trivial closure 
\begin{equation}
\langle  x c_{i} c_{n} \rangle=0. \label{mc}
\end{equation}
After using \eqref{mc} in \eqref{gi11}, the mean protein level can be computed exactly by solving a linear dynamical system given by \eqref{x3} and \eqref{gij}. At steady-state, the linear equations can be solved recursively to yield
\begin{equation}
\begin{aligned}
\overline{ \langle x c_{i} \rangle}= 
 \frac{\langle B \rangle}{\lambda_i}\frac{\sum_{j=1}^n \frac{k_j}{ \lambda_j}+\sum_{j=1}^i \frac{k_j}{ \lambda_j}}{\sum_{j=1}^n \frac{1}{\lambda_j}}, \label{all gene}
\end{aligned} 
\end{equation}
where $\langle B \rangle$ is the mean protein burst size. Since $c_i$'s are binary random variables, the mean protein level conditioned on the cell-cycle stage (i.e., synchronized cell population) can be obtained as
\begin{equation}
\begin{aligned}
\overline{ \langle x  \vert c_{i} \rangle}= \frac{\overline{ \langle x c_{i} \rangle}}{\overline{ \langle  c_{i} \rangle}}=
\langle B \rangle\left({\sum_{j=1}^n \frac{k_j}{ \lambda_j}+\sum_{j=1}^i \frac{k_j}{ \lambda_j}}\right). \label{all gene0}
\end{aligned} 
\end{equation}
Furthermore, using \eqref{all gene} and the fact that $\sum_{i=1}^n c_i=1$,
\begin{equation}
\overline{\langle x \rangle} = \sum_{j=1}^n \overline{\langle x c_{i} \rangle}=\frac{\langle B \rangle}{\sum_{j=1}^n \frac{1}{\lambda_j}}\left(
{\sum_{i=1}^n  \sum_{j=1}^n\frac{ k_j }{\lambda_i\lambda_j}}+{\sum_{i=1}^n  \sum_{j=1}^i\frac{ k_j }{\lambda_i\lambda_j}}\right).\label{mean0}
\end{equation}

Next, we investigate the mean protein level $\overline{\langle x \rangle}$ in some limiting cases. Consider equal transition rates between cell-cycle stages $\lambda_i=n/T$, which corresponds to an Erlang distributed cell-cycle durations with mean $T$ and shape parameter $n$. In this scenario
\begin{equation}
\overline{\langle x \rangle} =  \frac{ \langle B \rangle T\left(
\sum_{i=1}^n  \sum_{j=1}^n  k_j +
 \sum_{i=1}^n  \sum_{j=1}^i k_j\right)}{n^2},
\label{mean1}
\end{equation}
and further reduces to 
\begin{equation}
\overline{\langle x \rangle} = \langle B \rangle Tk \left(\frac{3}{2}+\frac{1}{2n}\right) \label{mean2}
\end{equation}
when  the rate of expression bursts $k_i=k$ is constant throughout the cell cycle. Finally, in the limit of deterministic cell-cycle durations of length $T$ ($n \to \infty$) 
\begin{equation}
\overline{\langle x \rangle} =\frac{ 3\langle B \rangle Tk}{2}  \label{mean3}.
\end{equation}

\section{Protein noise level for cell-cycle driven expression}
The mathematical approach illustrated above is now used to obtain the noise in protein copy numbers. By noise, we mean the magnitude of fluctuations in $x(t)$ that can be attributed to two stochastic mechanisms: bursty expression and random partitioning. Note that even in the absence of these mechanisms, there will be cell-cycle related fluctuations with protein molecules accumulating over time and dividing by half at random cell-division times. To correct for such cell-cycle driven fluctuations, we define another stochastic process $y(t)$ that estimates the protein level if expression and partitioning were modeled deterministically. More specifically, within the cell cycle $y(t)$ evolves according to the following differential equation
\begin{equation}
\dot{y}=\langle B \rangle\sum_{i=1}^n k_i  c_i(t) , \label{prody}
\end{equation}
which is the deterministic counterpart to the stochastic expression model presented earlier. At the time of cell division, the level is divided exactly by half
\begin{equation}
 y(t) \to  \frac{y (t) }{2}  \label{division character2}
\end{equation}
with zero partitioning errors, i.e., $\alpha=0$ in \eqref{division character}. This allows us to define a new zero-mean stochastic process $z(t)$ corrected for cell-cycle effects
\begin{equation}
z(t):= x(t)-y(t)  \label{division character3}
\end{equation}
that measures the deviation in the protein count in the original stochastic model ($x$) from its expected levels if noise mechanisms were modeled deterministically ($y$). The protein noise level can now be defined through the dimensionless quantify
\begin{equation}
CV^2 := \frac{\overline{\langle z^2 \rangle} }{\overline{\langle x \rangle}^2 },
\label{division character4}
\end{equation}
measuring the steady-state variance in $z(t)$ normalized by the square of the mean level. Since $\overline{\langle x \rangle}=\overline{\langle y \rangle}$ and $\overline{\langle xy \rangle}=\overline{\langle y^2 \rangle}$ (see Appendix B in SI), it can be rewritten as 
\begin{equation}
CV^2=\frac{\overline{\langle \left(x-y\right)^2 \rangle} }{\overline{\langle x \rangle}^2 }= \frac{\overline{\langle x^2 \rangle} }{\overline{\langle x \rangle}^2 }- \frac{\overline{\langle y^2 \rangle} }{\overline{\langle y \rangle}^2 }.
\label{division character4}
\end{equation}
In the context of prior work, $ {\overline{\langle y^2 \rangle} }/{\overline{\langle y \rangle}^2 }$ is interpreted as the ``extrinsic noise" in gene expression resulting from cell-cycle effects. It is typically measured by the covariance in the singe-cell expression of two identical copies of a gene with common cell-cycle regulation \cite{ses02,sis13}. In contrast, $CV^2$ is the  ``intrinsic noise"  resulting from stochasticity in gene expression and partitioning processes, and is measured by subtracting the extrinsic noise from the total noise $ {\overline{\langle x^2 \rangle} }/{\overline{\langle x \rangle}^2 }$. 

Having appropriately defined the noise level, we next compute it using moment equations.  The time evolution of the moments $\langle z^2 \rangle$ and $\langle z^2 c_i\rangle$
are given by (see Appendix C in SI)
\begin{subequations}\label{z2}
\begin{align}
&\frac{d\langle z^2 \rangle}{dt}= \langle B^2 \rangle \sum_{i=1}^n k_i \langle c_i \rangle +\frac{\alpha \lambda_n}{4} \langle xc_n\rangle +\frac{\lambda_n}{4}  \left \langle z^2c_1c_{n} \right \rangle - \frac{3}{4}\lambda_n \langle z^2 c_{n} \rangle\\
&\frac{d\langle z^2 c_{1} \rangle}{dt}=  k_1 \langle B^2 \rangle +\frac{\alpha \lambda_n}{4} \left \langle xc_{n} \right \rangle   +  \frac{\lambda_n}{4}  \left \langle z^2c_{n} \right \rangle - \lambda_{1} \langle z^2c_{1} \rangle, \label{x2g1}\\
&\frac{d\langle z^2 c_{i} \rangle}{dt}= k_i \langle B^2 \rangle  
- \lambda_{i} \langle z^2 c_{i} \rangle +\lambda_{i-1} \langle z^2 c_{i-1} \rangle, \ i=\left\lbrace 2,\ldots,i \right\rbrace. \label{x2gi}
\end{align} 
\end{subequations}
and depend on the fourth-order moments $\langle z^2c_1c_{n}\rangle$. Exploiting the model structure as before, it follows from \eqref{c} that $\langle z^2c_1c_{n}=0\rangle$, and    \eqref{x3}, \eqref{gij}, \eqref{z2} constitute a ``closed"  set linear differential equations. Steady-state analysis yields the following noise level (see Appendix C in SI)
\begin{align}
&CV^2= \underbrace{\left( \frac{1}{3}+\frac{2}{3}\frac{1}{1+\beta } \right)\frac{\langle B^2 \rangle }{\langle B \rangle }\frac{1 }{\overline{\langle x \rangle} }}_{\text{\normalsize Bursty synthesis}}+ \underbrace{ \frac{2 \alpha }{3}\frac{\beta }{1+\beta }\frac{1 }{\overline{\langle x \rangle} }}_{\text{\normalsize Partitioning errors}}\label{cv_division_time0}
\end{align}
that is inversely proportional to the mean $\overline{\langle x \rangle}$. The noise can be decomposed into two terms: the first term represents the contribution from protein synthesis in random bursts and depends on the statistical moments of the burst size $B$. The second term is the contribution from partitioning errors and depends linearly on $\alpha$. Recall that $\alpha$ measures the degree of randomness in partitioning of molecules between daughter cells, and is defined through \eqref{division character}. Interestingly, results show that the effect of cell-cycle regulation on the noise level can be quantified through a single dimensionless parameter
\begin{equation}
\beta = \frac{\sum_{i=1}^n  \sum_{j=1}^n\frac{ k_j }{\lambda_i\lambda_j}}{\sum_{i=1}^n  \sum_{j=1}^i\frac{ k_j }{\lambda_i\lambda_j}}, \label{beta}
\end{equation}
that is uniquely determined by the number of cell-cycle stages in the model ($n$), transition rates between stages ($\lambda_i$), and protein synthesis rates across stages ($k_i$). Note from \eqref{cv_division_time0} that $\beta$ affects the noise terms in opposite ways -- any coupling of cell-cycle to expression that increases $\beta$ will attenuate the contribution from bursty expression but amplifies the contribution from partitioning errors. Finally, we point out that in the case of non-bursty expression ($B=1$ with probability one) and binomial partitioning ($\alpha=1$)
\begin{align}
&CV^2=\frac{1}{\overline{\langle x \rangle}}.
\end{align}
and the noise level is always consistent with that of a Poisson distribution\footnote{The coefficient of variation squared for a Poisson distributed random variable is inverse of its mean} irrespective of the value of $\beta$, and hence the form of cell-cycle regulation.

\section{Optimal cell-cycle regulation to minimize noise}

We explore how different forms of cell-cycle regulation affect $CV^2$ and begin with the simplest case of
a constant synthesis rate  $k_i=k$, $i\in\{1,2,\ldots,n\}$ throughout the cell cycle. This case would correspond to a scenario where the net rate of expression (across all copies of a gene) remains invariant to replication-associated changes in gene dosage, as has recently been shown in different organisms \cite{pnb15,sxn16}. Further assuming equal transition rates $\lambda_i=n/T$ (Erlang distributed cell-cycle durations) 
\begin{equation}
\beta = \frac{2n}{n+1},
\end{equation}
which reduces to $\beta = 2$ as $n \to \infty$. Thus, in this important limit of no cell-cycle regulation (equal $k_i$'s) and deterministic cell-cycle duration (large $n$), 
\begin{align}
&CV^2=\frac{5}{9}\frac{\langle B^2 \rangle }{\langle B \rangle}\frac{1}{\overline{\langle x \rangle}}+ \frac{4 \alpha }{9}\frac{1}{\overline{\langle x \rangle}} \quad {\rm for} \ \ \beta=2.\label{cv_division_time000}
\end{align}
Next, consider the following strategies for coupling cell cycle to gene expression:
\begin{enumerate}
\item The burst arrival rate is assumed to increase by two-fold at the cell-cycle midpoint due to gene duplication. Assuming even $n$, this corresponds to 
\begin{subequations}
\begin{align}
k_i&=k, \ \  i\in \left\{1,\ldots, \frac{n}{2}\right\} \\
k_{i}&=2k, \ \  i\in \left\{\frac{n}{2},\ldots, n\right\} 
\end{align}
\end{subequations}
\item Expression only occurs at the start of cell cycle, i.e., $k_1=k$ and all other $k_i$'s are zero.
\item Expression only occurs at the end of cell cycle, i.e., $k_n=k$ and all other $k_i$'s are zero.
\item Expression only occurs at the cell cycle midpoint, i.e., $k_{\frac{n}{2}}=k$ and all other $k_i$'s are zero.
\end{enumerate}
For a mathematically controlled comparison, the parameter $k$ is adjusted using \eqref{mean0} from case-to-case so as to maintain a fixed average number protein molecules. The noise levels corresponding to the different forms of cell-cycle regulation are illustrated in Fig. 2. Interestingly, duplication of the protein expression rate within the cell cycle 
leads to a lower noise contribution from bursty synthesis, as compared to a constant rate throughout the cell-cycle. Moreover, expressing the protein only at the start (end) of cell cycle yields the highest (lowest) noise contribution from bursty synthesis. As expected from \eqref{cv_division_time0}, the noise contribution from partitioning errors exhibits a completely opposite trend (Fig. 2). 

\begin{figure}[!thb]
\centering
\includegraphics[width=0.75\linewidth]{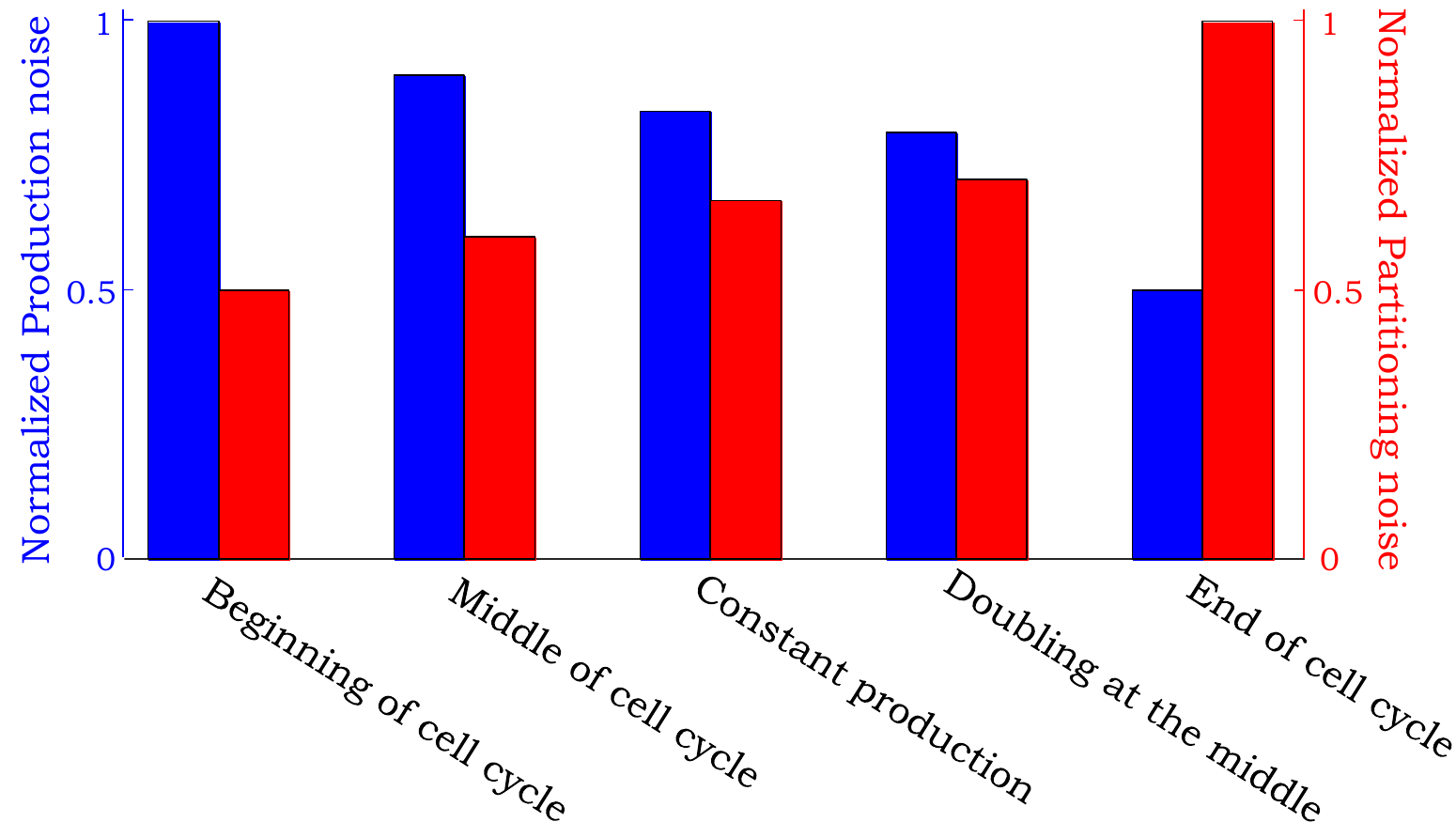}
\caption{{\bf Noise comparison for different strategies coupling cell cycle to gene expression}. The noise contributions from bursty expression (left) and partitioning errors (right) as given by \eqref{cv_division_time0} are shown for five different strategies: expression only at the start of cell cycle; expression only at the cell-cycle midpoint; constant mRNA synthesis rate throughout the cell cycle; doubling of synthesis rate at the cell-cycle midpoint; expression only towards the end of cell cycle. While noise contribution from bursty expression is minimized in the latter strategy, contribution from partitioning errors are lowest if expression occurs only at the beginning of cell cycle. The cell cycle was modeled by choosing $n=20$ stages with equal transition rates and production rates $k_i$ were chosen so as to have the same mean protein level per cell across all cases.}
\label{fig3}
\end{figure}

The above analysis begs an intriguing question: Is there an optimal way to express a protein during the cell cycle that maximizes/minimizes noise levels? Since the form of cell-cycle regulation impacts $CV^2$ through $\beta$, this amounts to choosing $k_i$'s so as to maximize/minimize it. Our result show that $\beta$ is bounded from both below and above (see Appendix D in SI)
\begin{equation}
1\leq \beta \leq \beta_{max}=\frac{\frac{1}{\lambda_1}+\frac{1}{\lambda_2}+\ldots+\frac{1}{\lambda_n}}{\frac{1}{\lambda_n}}. \label{max}
\end{equation}
The minimal value of $\beta=1$ is attained when expression only occurs at the start of cell cycle, i.e., a non-zero $k_1$ and all other $k_i$'s are zero. In this case
\begin{align}
&CV^2=\frac{2}{3}\frac{\langle B^2 \rangle }{\langle B \rangle}\frac{1}{\overline{\langle x \rangle}}+ \frac{\alpha }{3}\frac{1}{\overline{\langle x \rangle}} \quad {\rm for} \ \ \beta=1.\label{cv_division_time001}
\end{align}
with the lowest noise contribution from partitioning errors, but the highest contribution from bursty synthesis. In contrast, the maximum value of $\beta =\beta_{max}$ is attained when expression only occurs at the end of cell cycle, i.e., a non-zero $k_n$ and all other $k_i$'s are zero. Note form \eqref{max} that $\beta_{max} \to \infty$ as $\lambda_n \to \infty$ (time spent in stage $C_n$ approaches zero), in which case
\begin{align}
&CV^2=\frac{1}{3}\frac{\langle B^2 \rangle }{\langle B \rangle}\frac{1}{\overline{\langle x \rangle}}+ \frac{2\alpha }{3}\frac{1}{\overline{\langle x \rangle}} \quad {\rm for} \ \ \beta=\infty.\label{cv_division_time0001}
\end{align}
and the noise contribution from bursty synthesis is minimal. 

In summary, if bursty expression is the dominant source of noise (high $B$ and low $\alpha$), then $CV^2$ in minimized for a given $\overline{\langle x \rangle}$ when the protein is made in the shortest time window just before cell division (Fig. 3). On the other hand, if randomness in partitioning error is dominant (low $B$ and high $\alpha$), the optimal strategy is to make the protein just after cell division. Finally, we point out that these optimal strategies also minimize stochastic variation in protein counts among synchronized cells, where all cells are in the same cell-cycle stage (see Appendix E in SI).

\begin{figure}[!t]
\centering
\includegraphics[width=0.9\linewidth]{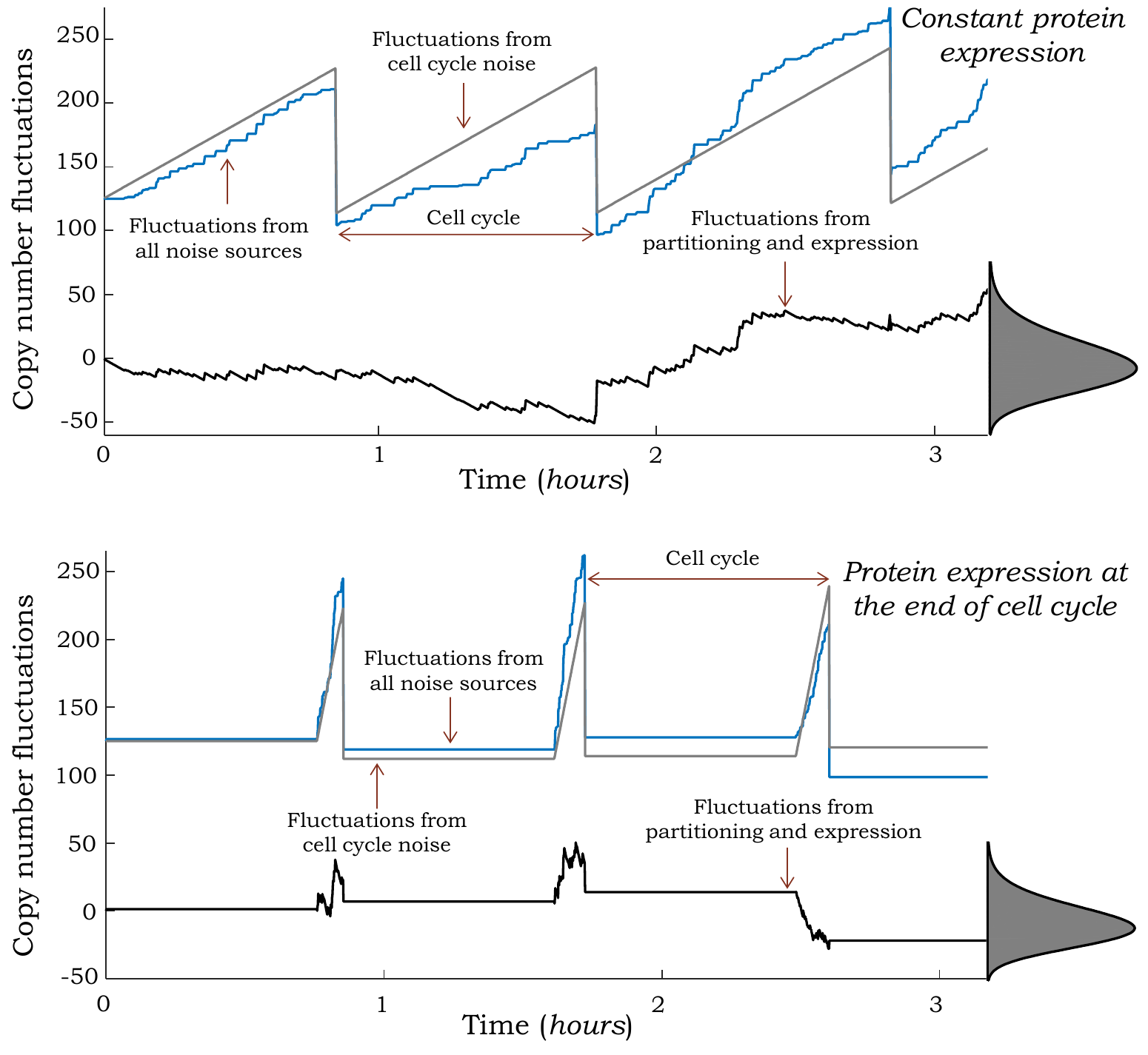}
\caption{{\bf Synthesis of proteins towards the end of cell cycle minimizes fluctuations in copy numbers}. Protein level in an individual cell across multiple cell cycles for two strategies: a fixed transcription rate throughout the cell cycle (top) and transcription only occurring just before cell division (bottom).
% In both cases, the spike train shows firings of burst events, t
Trajectories obtained via Monte Carlo simulations are shown for the stochastic model (blue) and a reduced model where noise mechanisms are modeled deterministically (gray). These levels are subtracted to obtain a zero-mean stochastic process $z(t)$, where fluctuations resulting from cell cycle are removed (black). Steady-state distribution of $z$ obtained from $10,000$ MC simulation runs is shown on the right, and the bottom strategy leads to lower variability in $z$ for the same mean protein level. Cell cycle and expression was modeled as in Fig. 1 and burst arrival rates were chosen so as to ensure a average protein copy number of $150$ molecules per cell in both cases.}
\label{fig2}
\end{figure}

\section{Discussion}
Theoretical model of stochastic gene expression have played a pivotal role in understanding how noise mechanisms and biologically relevant parameters generate differences in protein/mRNA population counts between isogenic cells \cite{tha16,boe13,kmk09,rak08,kis11,pau05,sbf15}. Here we have expanded this theory to consider cell-cycle regulated genes. Our approach involves a general model of cell cycle, where a cell transitioning through an arbitrary number of stages from birth to division. The protein is assumed to be expressed in random bursts, and the rate at which bursts arrive varies arbitrarily with cell-cycle stage. In the case of translational bursting of proteins from mRNA, the burst arrive rate corresponds to the mRNA synthesis (transcription) rate. In contrast, for transcriptional bursting of mRNAs, the burst arrive rate corresponds to the frequency with which a promoter become transcriptionally active. The key contribution of this work is derivation of \eqref{mean0} and \eqref{cv_division_time0} that predict the protein mean and noise levels for a given form of cell-cycle regulation.

Derivation of noise formulas enable uncovering of optimal cell-cycle regulation strategies to minimize $CV^2$ for a fixed mean protein level. In the physiological case of large bursts ($B \gg 1$) and binomial partitioning of proteins between daughter cells ($\alpha=1$), the contribution from bursty synthesis dominates $CV^2$. Our results show that in this scenario, expression of the protein just before division is the optimal strategy (Fig. 3). Intuitively, such a strategy can be understood in the context of the number of burst events from birth to division needed to maintain a given $\overline{\langle x \rangle}$ throughout the cell-cycle. It turns out that this number is highly dependent on the form of cell-cycle regulation. Hence, any strategy that requires more burst events to maintain the same mean protein level, lowers noise through more effective averaging of the underlying bursty process, albeit being more energy inefficient. For example, if protein production only occurs at the end of cell cycle, then on average, $\overline{\langle x \rangle}$ number of proteins have to added just before cell division. This corresponds to $\overline{\langle x \rangle}/\langle B \rangle$ number of burst events per cell cycle. If proteins were only expressed at the start of cell cycle, then one needs to add only $\overline{\langle x \rangle}/2$ number of molecules, half as much as the earlier strategy. If proteins were made at a constant synthesis rate throughout the cell cycle, then on average, $2\overline{\langle x \rangle}/3$ number of protein are added per cell cycle, which is higher than the early-expression strategy but lower than the late-expression strategy.  In summary, gene product synthesis just before division requires production of the most number of protein molecules to maintain a fixed mean level within the cell-cycle, and hence, provides the most effective noise buffering through averaging of burst events. Next, we provide two recent examples of proteins that are indeed expressed in this fashion.

The green alga \emph {C. reinhardtii} has a prolonged $G_1$ phase, where the size of a newborn cell increases by more than 2-fold. This long $G_1$ phase is followed by an $S/M$ phase. Here the cell undergoes multiple DNA replication and fission cycles creating $2^d$ daughter cells, where $d$ is number of rounds of division. Recent studies suggest that the number of rounds of division is controlled by a protein CDKG1, that is only expressed just before exit from $G_1$ \cite{lll16}. Another example, is the protein Whi5 in budding yeast \emph {S. cerevisiae} and its level controls the transition of cells past the Start checkpoint. This protein in not expressed in $G_1$, and is only synthesizes late in the  cell cycle \cite{stk15,scs15}. While such selective expression of these proteins plays a critical role in coupling cell size to cell-cycle decision, it may also minimize intrinsic fluctuations in protein levels from the innate stochasticity in gene expression. Clearly, a more systematic study exploring the role of noisy expression on the fidelity of these cell-cyle decisions is warranted.

%\subsection{Examples of cell-cycle regulated genes}

%\subsection{Future work}

It is important to point out that our analysis made various simplifying assumptions, such as, i) Excluding time evolution of cell size and size-dependent expression; ii) Instantaneous transcriptional and translational bursts that correspond to short-lived mRNAs and active promoter states;  iii) Cell-cycle durations being independent random variables, implying no correlation between the division times of mother and daughter cells. While many of these assumptions are clearly violated for cellular systems, they were necessary to obtain exact analytical solutions that provide novel insights into noise control by synchronizing gene expression to cell cycle. Further work will focus on relaxing these assumptions, in particular, the first assumption of incorporating cell size into the model. This will allow investigation of both concentration and copy number of gene products in single cells, and some recent work on modeling stochastic dynamics of cell size has already been done \cite{ama14,gvs15,vas16}.

\section*{Author contributions}
AS defined the problem and formulated the approach. MS did the mathematical derivations and both authors collaborated on writing the paper.

\section*{Acknowledgments}
AS is supported by the National Science Foundation Grant DMS-1312926. We thank Cesar Vargas-Garcia and Khem Ghusinga for feedback on the manuscript.

\appendix

\section*{Appendix}

\section{Moment equations describing the model}

Based on standard stochastic formulation of chemical kinetics \cite{mcq67,gil01}, the model describing $x$ contains the following stochastic events 
\begin{subequations}
\begin{align}
& \text{Protein production: }&&x\xmapsto{k_ic_ip_j}x+j,\\
& \text{Cell stage evolution: }&&c_i\xmapsto{\lambda_ic_i}c_i -1, \ \ c_{i+1}\xmapsto{\lambda_ic_i}c_{i+1}+1, \\ 
& \text{Cell division: }&&x\xmapsto{\lambda_n c_n}x_+, \ \ c_n\xmapsto{\lambda_nc_n}c_n-1, \ \ c_{1}\xmapsto{\lambda_nc_n}c_{1}+1, 
\end{align}
\label{reaction x}
\end{subequations}
where the probability of having a burst of $j$ molecules is given by $p_j$. Whenever an event occurs, the states of the system change based on the stochiometries given in \eqref{reaction x}. On top of the arrows we showed the event propensity function $\psi (x,c_{i})$, which determines how often reactions occur, i.e., the probability that an event occurs in the next infinitesimal time interval $(t,t+dt]$ is $\psi(x,c_{i})dt$. Time derivative of the expected value of any function $\varphi(x,c_{i})$ for this system can be written as \cite{hsi04}
\begin{equation}
\begin{aligned}
\frac{d\langle \varphi(x,c_{i}) \rangle}{dt}=& \left \langle \sum_{Events}  \Delta \varphi(x,c_{i}) \times \psi(x,c_{i})\right \rangle .   
\label{dynnf}
\end{aligned}
\end{equation}
Choosing $\varphi$ to be $x$ and $c_i, i=\{1,2,\ldots,n\}$ results in the equation (6) in the main article.

\section{Moment dynamics of $y$}
The model describing $x$ and $y$ includes the stochastic events
\begin{subequations}
\begin{align}
& \text{Protein production: }&&x\xmapsto{k_ic_ip_j}x+j,\\
& \text{Cell stage evolution: }&&c_i\xmapsto{\lambda_ic_i}c_i -1, \ c_{i+1}\xmapsto{\lambda_ic_i}c_{i+1}+1, \\ 
& \text{Cell division: }&&x\xmapsto{\lambda_n c_n}x_+,\ y\xmapsto{\lambda_n c_n}y_+, \ c_n\xmapsto{\lambda_nc_n}c_n-1, \ c_{1}\xmapsto{\lambda_nc_n}c_{1}+1, 
\end{align}
\label{reaction x and y}
\end{subequations}
and the deterministic production of $y$
\begin{equation}
\dot{y}= \left(\sum_{i=1}^n k_i  c_i\right) \langle B \rangle. \label{prod. y}
\end{equation}
Time derivative of the expected value of any function $\varphi(x,y,c_{i})$ for this system can be written as \cite{hsi04}
\begin{equation}
\begin{aligned}
\frac{d\langle \varphi(x,y,c_{i}) \rangle}{dt}=& \left \langle \sum_{Events}  \Delta \varphi(x,y,c_{i}) \times \psi(x,y,c_{i})\right \rangle + \left \langle\left(\sum_{i=1}^n k_i  c_i\right) \frac{\partial \varphi(x,y,c_{i})}{\partial z}  \langle B \rangle \right \rangle ,   
\label{dynnf0}
\end{aligned}
\end{equation}
where the first term in the right-hand side is contributed from stochastic events and the second one is contributed from \eqref{prod. y}. The propensity function of events is given by $\psi (x,y,c_{i})$. 
The mean dynamics of $y$ can be written by choosing $\varphi$ to be $y$ 
\begin{equation}
\frac{d\langle y \rangle}{dt}= \left(\sum_{i=1}^n k_i \langle c_i \rangle\right)\langle B \rangle - \frac{\lambda_n}{2} \langle y c_{n} \rangle. \label{x0}
\end{equation}
Dynamics of $\langle y \rangle$ is not closed and depends to moments $\langle yc_{n} \rangle $, hence in order to have a closed set of equations we add new moments dynamics by selecting $\varphi$ to be $ yc_{i} $
\begin{subequations}
\begin{align}
&  \frac{d\langle y c_{1} \rangle}{dt}=k_1 \langle B \rangle  \langle c_{1} \rangle + \frac{\lambda_n}{2}\left \langle y c_{n} \right\rangle    - \lambda_1 \langle yc_{1} \rangle,\\
&\frac{d\langle y c_{i} \rangle}{dt}=  k_i \langle B \rangle  \langle c_{i} \rangle 
- \lambda_{i} \langle y c_{i} \rangle +\lambda_{i-1}\langle y c_{i-1} \rangle , \ \ j\in \lbrace 2,\ldots,n \rbrace.\label{gij tot0}
\end{align} 
\label{gij0}
\end{subequations}
Dynamics of $\langle y \rangle $ and $\langle yc_i \rangle, \ j\in \lbrace 1,\ldots,n \rbrace$ are the same as dynamics of  $\langle x \rangle $ and $\langle xc_i \rangle, \ j\in \lbrace 1,\ldots,n \rbrace$ presented in (6b) and (8) in the main text, hence $\overline{\langle x \rangle} =\overline{\langle y \rangle} $ and $\overline{\langle xc_i \rangle} =\overline{\langle yc_i \rangle} $.

Further, dynamics of $\langle x y \rangle $ can be written as
\begin{align}
&\frac{d\langle x y  \rangle}{dt}= \left(\sum_{i=1}^n k_i \left(\langle xc_i \rangle + \langle yc_i \rangle \right)\right)\langle B \rangle - \frac{\lambda_n}{4} \langle x yc_{n} \rangle. \label{xy}
\end{align} 
In order to have a closed set of equations we add dynamics of $ \langle x yc_{i} \rangle$
\begin{subequations}
\begin{align}
&\frac{d\langle xy c_{1} \rangle}{dt}= k_1 \left(\langle x c_1\rangle+ \langle yc_1 \rangle \right)\langle B \rangle  +  \frac{\lambda_n}{4}  \left \langle xyc_{n} \right \rangle - \lambda_{1} \langle xyc_{1} \rangle, \label{xyg1}\\
&\frac{d\langle xy c_{i} \rangle}{dt}=  k_i \left(\langle x c_i\rangle+ \langle yc_i \rangle \right) \langle B \rangle
- \lambda_{i} \langle xy c_{i} \rangle +\lambda_{i-1} \langle xy c_{i-1} \rangle, \ i=\left\lbrace 2,\ldots,i \right\rbrace. \label{xygi}
\end{align} 
\label{xyg}
\end{subequations}
By having a closed set of equations related to $xy$, in the next step we add dynamics of $\langle y^2 \rangle $ and $\langle y^2 c_i \rangle $
\begin{subequations}
\begin{align}
&\frac{d\langle y^2  \rangle}{dt}= 2 \left(\sum_{i=1}^n k_i  \langle yc_i \rangle \right)\langle B \rangle - \frac{\lambda_n}{4} \langle y^2c_{n} \rangle, \label{y2}\\
&\frac{d\langle y^2 c_{1} \rangle}{dt}= 2 k_1  \langle yc_1 \rangle \langle B \rangle  +  \frac{\lambda_n}{4}  \left \langle y^2c_{n} \right \rangle - \lambda_{1} \langle y^2c_{1} \rangle, \label{y2g1}\\
&\frac{d\langle y^2 c_{i} \rangle}{dt}= 2 k_i \langle y c_i\rangle \langle B \rangle
- \lambda_{i} \langle y^2 c_{i} \rangle +\lambda_{i-1} \langle y^2 c_{i-1} \rangle, \ i=\left\lbrace 2,\ldots,i \right\rbrace. \label{y2gi}
\end{align} 
\label{all y}
\end{subequations}
Using the fact that $\overline{\langle x \rangle} =\overline{\langle y \rangle} $ and $\overline{\langle xc_i \rangle} =\overline{\langle yc_i \rangle} $, equations  \eqref{all y}, \eqref{xy}, and \eqref{xyg} in steady-state results in $\overline{\langle y^2 \rangle} =\overline{\langle xy \rangle} $ and $\overline{\langle y^2c_i \rangle} =\overline{\langle xyc_i \rangle} $.

\section{Calculation of $z^2$}
The random variable $z$ is governed via  
\begin{subequations}
\begin{align}
&z(t) \mapsto z(t)+ B , \\ &\dot{z}=- \left(\sum_{i=1}^n k_i  c_i\right) \langle B \rangle\label{modelz} .
\end{align}
\label{whole model of z}
\end{subequations}
Further in the time of division, $z_+$ is defined as
\begin{equation}
  \langle z_+(t_s) \vert  z(t_s)   \rangle= \frac{z(t_s) }{2},  \ \ 
\left \langle{z_+^2(t_s)} -\langle z_+(t_s) \rangle^2\bigg\vert  z(t_s) \right \rangle = \frac{ \alpha x(t_s)}{4}. \label{w character}
\end{equation}

%\begin{figure}[h]
%\centering
%\includegraphics[width=\textwidth]{table1.pdf} 
%\end{figure}

Hence the model by taking into account $z$ contains the following stochastic events
\begin{subequations}
\begin{align}
& \text{Protein production: }&&x\xmapsto{k_ic_ip_j}x+j,\  z\xmapsto{k_ic_ip_j}z+j, \\
& \text{Cell stage evolution: }&&c_i\xmapsto{\lambda_ic_i}c_i -1, \ c_{i+1}\xmapsto{\lambda_ic_i}c_{i+1}+1, \\ 
& \text{Cell division: }&&x\xmapsto{\lambda_n c_n}x_+,\ z\xmapsto{\lambda_n c_n}z_+,  \ c_n\xmapsto{\lambda_nc_n}c_n-1,  \ c_{1}\xmapsto{\lambda_nc_n}c_{1}+1, 
\end{align}
\label{reaction z}
\end{subequations}
and deterministic dynamics of $z$ given in \eqref{modelz}.
Time derivative of the expected value of any function $\varphi(x,z,c_{i})$ for this system can be written as \cite{hsi04}
\begin{equation}
\begin{aligned}
\frac{d\langle \varphi(x,z,c_{i}) \rangle}{dt}=& \left \langle \sum_{Events}  \Delta \varphi(x,z,c_{i}) \times \psi(x,z,c_{i})\right \rangle - \left \langle\left(\sum_{i=1}^n k_i  c_i\right) \frac{\partial \varphi(x,z,c_{i})}{\partial z}  \langle B \rangle \right \rangle ,   
\label{dynnf00}
\end{aligned}
\end{equation}
where the first term in the right-hand side is contributed from stochastic events and the second one is contributed from \eqref{modelz}. The propensity function of events is given by $\psi (x,z,c_{i})$.

By choosing $\varphi$ to be $z^2$ and $z^2 c_{i}, i=\left\lbrace 1,\ldots,i \right\rbrace$ we have the following moment dynamics
\begin{subequations}
\begin{align}
&\frac{d\langle z^2 \rangle}{dt}= \left(\sum_{i=1}^n k_i \langle c_i \rangle\right)\langle B^2 \rangle +\frac{1}{4}\alpha \lambda_n \langle xc_n\rangle - \frac{3}{4}\lambda_n \langle z^2 c_{n} \rangle,\label{z20}\\
&\frac{d\langle z^2 c_{1} \rangle}{dt}=  k_1 \langle B^2 \rangle +\frac{1}{4}\alpha \lambda_n \left \langle xc_{n} \right \rangle   +  \frac{\lambda_n}{4}  \left \langle z^2c_{n} \right \rangle - \lambda_{1} \langle z^2c_{1} \rangle, \label{z2g1}\\
&\frac{d\langle z^2 c_{i} \rangle}{dt}= k_i \langle B^2 \rangle  
- \lambda_{i} \langle z^2 c_{i} \rangle +\lambda_{i-1} \langle z^2 c_{(i-1)} \rangle, \ i=\left\lbrace 2,\ldots,i \right\rbrace. \label{z2gi}
\end{align} 
\end{subequations}
Note that just one of the binary states $c_{i}$ can be $1$ at a time, thus $\overline{ \langle z^2 \rangle} =\sum_{i=1}^{n}\overline{  \left\langle z^2 c_{i} \right\rangle} $. In order to calculate the terms $\overline{ \langle z^2 c_{i} \rangle}$ we need to express the term $\overline{ \left \langle   z^2 c_{n} \right \rangle}$ as the first step. This term can be calculated by analyzing equation \eqref{z20} in steady-state 
\begin{equation}
\begin{aligned}
\overline{ \left  \langle  z^2 c_{n} \right \rangle }= \frac{4}{3\lambda_n}\frac{\sum_{j=1}^n k_j \lambda_j}{\sum_{j=1}^n \lambda_j}\langle B^2 \rangle +\frac{2\alpha }{3 \lambda_n}\frac{\sum_{j=1}^n k_j \lambda_j}{\sum_{j=1}^n \lambda_j}\langle B \rangle  .\label{last gene0}
\end{aligned} 
\end{equation}
By using a recursive process we calculate moments $\overline{ \langle z^2 c_{i} \rangle}$: we calculate $\overline{ \langle z^2 c_{1} \rangle}$ by substituting equation \eqref{last gene0} in equation \eqref{z2g1}. Then we use the definition of $\overline{\langle z^2 c_{1} \rangle}$ to calculate $\overline{\langle z^2 c_{2} \rangle}$ from equation \eqref{z2gi} and so on
\begin{equation}
\begin{aligned}
& \overline{ \langle z^2 c_{i} \rangle}= \frac{1}{3\lambda_i}\frac{\sum_{j=1}^n k_j \lambda_j}{\sum_{j=1}^n \lambda_j}\langle B^2 \rangle  + \frac{1}{\lambda_i}\frac{\sum_{j=1}^i k_j \lambda_j}{\sum_{j=1}^n \lambda_j}\langle B^2 \rangle +\frac{2\alpha }{3 \lambda_i}\frac{\sum_{j=1}^n k_j \lambda_j}{\sum_{j=1}^n \lambda_j}\langle B \rangle .\label{all rgi}
\end{aligned} 
\end{equation}
Summing up all the term in equation \eqref{all rgi} results in $\overline{\langle z^2 \rangle} $
\begin{equation}
\overline{\langle z^2 \rangle}= \frac{1}{3\sum_{j=1}^n \lambda_j}\sum_{i=1}^n  \sum_{j=1}^n\frac{ k_j \lambda_j}{\lambda_i}\langle B^2 \rangle+
 \frac{1}{\sum_{j=1}^n \lambda_j}\sum_{i=1}^n  \sum_{j=1}^i\frac{ k_j \lambda_j}{\lambda_i}\langle B^2 \rangle +\frac{2 \alpha}{3\sum_{j=1}^n \lambda_j}\sum_{i=1}^n  \sum_{j=1}^n\frac{ k_j \lambda_j}{\lambda_i}\langle B \rangle .
\end{equation}
Finally, protein noise level can be written as
\begin{align}
&CV^2=  \frac{\overline{\langle z^2 \rangle}}{\overline{\langle x \rangle}^2}= \left( \frac{1}{3}+\frac{2}{3}\frac{1}{1+\beta } \right)\frac{\langle B^2 \rangle }{\langle B \rangle }\frac{1 }{\overline{\langle x \rangle} }+  \frac{2 \alpha }{3}\frac{\beta }{1+\beta }\frac{1 }{\overline{\langle x \rangle} },\label{cv_division_time00}
\end{align}
where 
\begin{equation}
\beta = \frac{\sum_{i=1}^n  \sum_{j=1}^n\frac{ k_j \lambda_j}{\lambda_i}}{\sum_{i=1}^n  \sum_{j=1}^i\frac{ k_j \lambda_j}{\lambda_i}}.
\end{equation}

\section{Optimal value of $\beta$}
From \eqref{cv_division_time00} it is clear that minimum production noise occurs when $\beta$ is maximum, and minimum value of partitioning noise happens when $\beta$ is minimum. $\beta$ can be written as
\begin{equation}
\beta = \frac{\sum_{i=1}^n  \sum_{j=1}^n\frac{ k_j \lambda_j}{\lambda_i}}{\sum_{i=1}^n  \sum_{j=1}^i\frac{ k_j \lambda_j}{\lambda_i}}=\frac{k_1\lambda_1 a_1+k_2\lambda_2 a_1+\ldots+ k_n\lambda_n a_1}{k_1\lambda_1 a_1+k_2\lambda_2 a_2+\ldots+ k_n\lambda_n a_n},
\end{equation}
where 
\begin{equation}
a_1 = \frac{1}{\lambda_1}+\frac{1}{\lambda_2}+\ldots+\frac{1}{\lambda_n}, \ \ a_2 = \frac{1}{\lambda_2}+\ldots+\frac{1}{\lambda_n}, \ \  a_n = \frac{1}{\lambda_n}.
\end{equation}
Note that 
\begin{equation}
a_1 >a_2>\ldots>a_n \Rightarrow \beta \leq \frac{a_1}{a_n},
\end{equation}
where equality happens when all $k_i$s are zero except $k_n$. Using the same methodology one can see that minimum of $\beta$ happens when all the rates are zero except $k_1$. The minimum value of $\beta$ is one.

\section{Noise in synchronized cells}
Statistical moments conditioned on the cell cycle stage $C_{i}$ can be obtained using
\begin{align}
&\overline{ \langle x \vert  c_{i} \rangle}=\frac{\overline{\langle x c_{i} \rangle}}{\overline{\langle c_{i} \rangle}}, \ \overline{ \langle x^2 \vert  c_{i} \rangle}=\frac{\overline{ \langle x^2 c_{i} \rangle}}{\overline{ \langle c_{i} \rangle}}  \label{cond}.
\end{align}
In order to calculate noise in synchronized cells we need to calculate $\langle x^2 c_{i} \rangle$
\begin{subequations}
\begin{align}
&\frac{d\langle x^2 c_{1} \rangle}{dt}= 2k_1\langle xc_1 \rangle+  k_1 \langle B^2 \rangle +\frac{1}{4}\alpha \lambda_n \left \langle xc_{n} \right \rangle   +  \frac{\lambda_n}{4}  \left \langle x^2c_{n} \right \rangle - \lambda_{1} \langle x^2c_{1} \rangle, \label{xx2g1}\\
&\frac{d\langle x^2 c_{i} \rangle}{dt}= 2k_i\langle xc_i \rangle + k_i \langle B^2 \rangle  
- \lambda_{i} \langle x^2 c_{i} \rangle +\lambda_{i-1} \langle x^2 c_{(i-1)} \rangle, \ i=\left\lbrace 2,\ldots,i \right\rbrace. \label{xx2gi}
\end{align} 
\end{subequations}
In order to calculate $\langle x^2 c_{n} \rangle$ we introduce the moment dynamics of $\langle x^2 \rangle$
\begin{equation}
\begin{aligned}
&\frac{d\langle x^2 \rangle}{dt}= 
2\left(\sum_{i=1}^n k_i \langle xc_i \rangle\right)\langle B \rangle +
\left(\sum_{i=1}^n k_i \langle c_i \rangle\right)\langle B^2 \rangle +\frac{1}{4}\alpha \lambda_n \langle xc_n\rangle - \frac{3}{4}\lambda_n \langle x^2 c_{n} \rangle,
\end{aligned} \label{xx2}
\end{equation}
hence in steady-state 
\begin{equation}
\begin{aligned}
\overline{ \left  \langle  x^2 c_{n} \right \rangle }=  \frac{8}{3\lambda_n}\frac{\left(\sum_{j=1}^n \frac{k_j}{ \lambda_j}\right)^2+\sum_{i=1}^n \frac{k_i}{ \lambda_i} \sum_{j=1}^i \frac{k_j}{ \lambda_j}}{\sum_{j=1}^n \frac{1}{\lambda_j}}\langle B \rangle + \frac{4}{3\lambda_n}\frac{\sum_{j=1}^n \frac{k_j}{ \lambda_j}}{\sum_{j=1}^n \frac{1}{\lambda_j}}\langle B^2 \rangle +\frac{2\alpha }{3 \lambda_n}\frac{\sum_{j=1}^n \frac{k_j}{ \lambda_j}}{\sum_{j=1}^n \frac{1}{\lambda_j}}\langle B \rangle  .\label{last gene00}
\end{aligned} 
\end{equation}
By using a similar process used in the previous section we calculate moments $\overline{ \langle x^2 c_{i} \rangle}$
\begin{equation}
\begin{aligned}
\overline{ \langle x^2 c_{i} \rangle}= \frac{2}{3\lambda_i}\frac{\left(\sum_{j=1}^n \frac{k_j}{ \lambda_j}\right)^2}{\sum_{j=1}^n \frac{1}{\lambda_j}}\langle B \rangle 
+
 \frac{2}{3\lambda_i}\frac{\sum_{i=1}^n\frac{k_i}{ \lambda_i} \sum_{j=1}^i \frac{k_j}{ \lambda_j}}{\sum_{j=1}^n \frac{1}{\lambda_j}}\langle B \rangle 
 + \frac{2}{\lambda_i}\frac{\sum_{s=1}^i \frac{k_s}{ \lambda_s} \sum_{j=1}^s \frac{k_j}{ \lambda_j}}{\sum_{j=1}^n \frac{1}{\lambda_j}}\langle B \rangle 
\\
 + \frac{2}{\lambda_i}\frac{ \sum_{j=1}^i \frac{k_j}{ \lambda_j}\sum_{i=1}^n \frac{k_i}{ \lambda_i}}{\sum_{j=1}^n \frac{1}{\lambda_j}}\langle B \rangle  + \frac{1}{3\lambda_i}\frac{\sum_{j=1}^n \frac{k_j}{ \lambda_j}}{\sum_{j=1}^n \frac{1}{\lambda_j}}\langle B^2 \rangle  + \frac{1}{\lambda_i}\frac{\sum_{j=1}^i  \frac{k_j}{ \lambda_j}}{\sum_{j=1}^n \frac{1}{\lambda_j}}\langle B^2 \rangle +\frac{2\alpha }{3 \lambda_i}\frac{\sum_{j=1}^n  \frac{k_j}{ \lambda_j}}{\sum_{j=1}^n \frac{1}{\lambda_j}}\langle B \rangle .\label{all xgi}
\end{aligned} 
\end{equation}
By having $\overline{ \langle xc_i \rangle} $ and $\overline{ \langle x^2c_i \rangle} $ from (10) and \eqref{all xgi}, we can calculate mean and noise in synchronized cells. 
Using \eqref{cond} yields the following conditional mean 
\begin{equation}
\begin{aligned}
\langle x \vert {c_{i}=1} \rangle=  \left(\sum_{j=1}^n \frac{k_j}{ \lambda_j}+\sum_{j=1}^i \frac{k_j}{ \lambda_j}\right)\langle B \rangle.
\end{aligned} \label{sync-mean}
\end{equation}
Further, the protein noise level given that cells are in stage $C_{i}$ is given by 
\begin{equation}
\begin{aligned}
 CV^2\vert_{c_{i}=1} =&
\underbrace{\left(\frac{2}{3}\sum_{j=1}^n \left(\frac{1}{\lambda_j}\right)\frac{\sum_{i=1}^n\frac{k_i}{ \lambda_i} \sum_{j=1}^i \frac{k_j}{ \lambda_j}+3\sum_{s=1}^i \frac{k_s}{ \lambda_s} \sum_{j=1}^s \frac{k_j}{ \lambda_j}+\left( \sum_{j=1}^i \frac{k_j}{ \lambda_j}\right)^2(3+\beta_c )\beta_c}{\left( \sum_{j=1}^i \frac{k_j}{ \lambda_j}\right)^2(1+\beta_c )^2}-1\right)}_{\text{Cell cycle variations}}\\
\\
&+
 \underbrace{\left( \frac{1}{3}+\frac{2}{3}\frac{1}{1+\beta_c } \right)\frac{\langle B^2 \rangle }{\langle B \rangle }\frac{1}{ \langle x\vert c_{i} \rangle}}_{\text{Burst synthesis noise}} + \underbrace{\frac{2 \alpha }{3}\frac{\beta_c}{1+\beta_c }\frac{1}{ \langle x\vert c_{i} \rangle}}_{\text{Partitioning errors}},
\end{aligned} 
\end{equation}
where
\begin{equation}
\beta_c = \frac{ \sum_{j=1}^n \frac{k_j}{ \lambda_j}}{ \sum_{j=1}^i \frac{k_j}{ \lambda_j}}.
\end{equation}

\bibliographystyle{IEEEtran}
\bibliography{references}

\begin{thebibliography}{10}
\providecommand{\url}[1]{#1}
\csname url@samestyle\endcsname
\providecommand{\newblock}{\relax}
\providecommand{\bibinfo}[2]{#2}
\providecommand{\BIBentrySTDinterwordspacing}{\spaceskip=0pt\relax}
\providecommand{\BIBentryALTinterwordstretchfactor}{4}
\providecommand{\BIBentryALTinterwordspacing}{\spaceskip=\fontdimen2\font plus
\BIBentryALTinterwordstretchfactor\fontdimen3\font minus
  \fontdimen4\font\relax}
\providecommand{\BIBforeignlanguage}[2]{{%
\expandafter\ifx\csname l@#1\endcsname\relax
\typeout{** WARNING: IEEEtran.bst: No hyphenation pattern has been}%
\typeout{** loaded for the language `#1'. Using the pattern for}%
\typeout{** the default language instead.}%
\else
\language=\csname l@#1\endcsname
\fi
#2}}
\providecommand{\BIBdecl}{\relax}
\BIBdecl

\bibitem{bkc03}
W.~J. Blake, M.~Kaern, C.~R. Cantor, and J.~J. Collins, ``Noise in eukaryotic
  gene expression,'' \emph{Nature}, vol. 422, pp. 633--637, 2003.

\bibitem{rao05}
J.~M. Raser and E.~K. O'Shea, ``Noise in gene expression: Origins,
  consequences, and control,'' \emph{Science}, vol. 309, pp. 2010 -- 2013,
  2005.

\bibitem{keb05}
M.~K{\ae}rn, T.~C. Elston, W.~J. Blake, and J.~J. Collins, ``Stochasticity in
  gene expression: from theories to phenotypes,'' \emph{{Nature Reviews
  Genetics}}, vol.~6, pp. 451--464, 2005.

\bibitem{nmt_13}
G.~Neuert, B.~Munsky, R.~Z. Tan, L.~Teytelman, M.~Khammash, and A.~v.
  Oudenaarden, ``Systematic identification of signal-activated stochastic gene
  regulation,'' \emph{Science}, vol. 339, pp. 584--587, 2013.

\bibitem{mal13}
A.~Magklara and S.~Lomvardas, ``Stochastic gene expression in mammals: lessons
  from olfaction,'' \emph{Trends in Cell Biology}, vol.~23, pp. 449--456, 2014.

\bibitem{jbp14}
M.~Jangi, P.~L. Boutz, P.~Paul, and P.~A. Sharp, ``Rbfox2 controls
  autoregulation in {RNA}-binding protein networks,'' \emph{Genes \&
  Development}, vol.~28, pp. 637--651, 2014.

\bibitem{sgz11}
L.-H. So, A.~Ghosh, C.~Zong, L.~A. Sep{\'u}lveda, R.~Segev, and I.~Golding,
  ``General properties of transcriptional time series in escherichia coli,''
  \emph{Nature genetics}, vol.~43, pp. 554--560, 2011.

\bibitem{smg11}
D.~M. Suter, N.~Molina, D.~Gatfield, K.~Schneider, U.~Schibler, and F.~Naef,
  ``Mammalian genes are transcribed with widely different bursting kinetics,''
  \emph{Science}, vol. 332, pp. 472--474, 2011.

\bibitem{drs12}
R.~D. Dar, B.~S. Razooky, A.~Singh, T.~V. Trimeloni, J.~M. McCollum, C.~D. Cox,
  M.~L. Simpson, and L.~S. Weinberger, ``Transcriptional burst frequency and
  burst size are equally modulated across the human genome,'' \emph{Proceedings
  of the National Academy of Sciences}, vol. 109, pp. 17\,454--17\,459, 2012.

\bibitem{singh_transient_2014}
A.~Singh, ``Transient changes in intercellular protein variability identify
  sources of noise in gene expression,'' \emph{Biophysical Journal}, vol. 107,
  pp. 2214--2220, 2014.

\bibitem{hbr12}
G.~Hornung, R.~Bar-Ziv, D.~Rosin, N.~Tokuriki, D.~S. Tawfik, M.~Oren, and
  N.~Barkai, ``Noise-mean relationship in mutated promoters,'' \emph{Genome
  Research}, vol.~22, pp. 2409--2417, 2012.

\bibitem{rpt06}
A.~Raj, C.~Peskin, D.~Tranchina, D.~Vargas, and S.~Tyagi, ``Stochastic m{RNA}
  synthesis in mammalian cells,'' \emph{PLOS Biology}, vol.~4, p. e309, 2006.

\bibitem{srd12}
A.~Singh, B.~S. Razooky, R.~D. Dar, and L.~S. Weinberger, ``Dynamics of protein
  noise can distinguish between alternate sources of gene-expression
  variability,'' \emph{Molecular Systems Biology}, vol.~8, p. 607, 2012.

\bibitem{btl14}
K.~B. Halpern, S.~Tanami, S.~Landen, M.~Chapal, L.~Szlak, A.~Hutzler,
  A.~Nizhberg, and S.~Itzkovitz, ``Bursty gene expression in the intact
  mammalian liver,'' \emph{Molecular Cell}, vol.~58, pp. 147--156, 2015.

\bibitem{bhh16}
C.~R. Bartman, S.~C. Hsu, C.~C.-S. Hsiung, A.~Raj, and G.~A. Blobel, ``Enhancer
  regulation of transcriptional bursting parameters revealed by forced
  chromatin looping,'' \emph{Molecular Cell}, vol.~62, pp. 237 -- 247, 2016.

\bibitem{dsp15}
B.~Daigle, M.~Soltani, L.~Petzold, and A.~Singh, ``Inferring single-cell gene
  expression mechanisms using stochastic simulation,'' \emph{Bioinformatics},
  vol.~31, pp. 1428--1435, 2015.

\bibitem{brb14}
C.~R. Brown and H.~Boeger, ``Nucleosomal promoter variation generates gene
  expression noise,'' \emph{Proceedings of the National Academy of Sciences},
  vol. 111, pp. 17\,893--17\,898, 2014.

\bibitem{cfx06}
L.~Cai and N.~F. X.~S. Xie, ``Stochastic protein expression in individual cells
  at the single molecule level,'' \emph{Nature}, vol. 440, pp. 358--362, Sep.
  2006.

\bibitem{shs08}
V.~Shahrezaei and P.~S. Swain, ``Analytical distributions for stochastic gene
  expression,'' \emph{Proceedings of the National Academy of Sciences}, vol.
  105, pp. 17\,256--17\,261, 2008.

\bibitem{sih09b}
A.~Singh and J.~P. Hespanha, ``Optimal feedback strength for noise suppression
  in autoregulatory gene networks,'' \emph{Biophysical Journal}, vol.~96, pp.
  4013--4023, 2009.

\bibitem{otk02}
E.~M. Ozbudak, M.~Thattai, I.~Kurtser, A.~D. Grossman, and A.~van Oudenaarden,
  ``Regulation of noise in the expression of a single gene,'' \emph{Nature
  Genetics}, vol.~31, pp. 69--73, 2002.

\bibitem{ccf08}
P.~J. Choi, L.~Cai, K.~Frieda, and X.~S. Xie, ``A stochastic single-molecule
  event triggers phenotype switching of a bacterial cell,'' \emph{Science},
  vol. 322, pp. 442--446, 2008.

\bibitem{src10}
A.~Singh, B.~Razooky, C.~D. Cox, M.~L. Simpson, and L.~S. Weinberger,
  ``Transcriptional bursting from the {HIV}-1 promoter is a significant source
  of stochastic noise in {HIV}-1 gene expression,'' \emph{Biophysical Journal},
  vol.~98, pp. {L}32--{L}34, 2010.

\bibitem{Eldar:2010kk}
A.~Eldar and M.~B. Elowitz, ``Functional roles for noise in genetic circuits,''
  \emph{Nature}, vol. 467, pp. 167--173, 2010.

\bibitem{siw09}
A.~Singh and L.~S. Weinberger, ``Stochastic gene expression as a molecular
  switch for viral latency,'' \emph{Current Opinion in Microbiology}, vol.~12,
  pp. 460--466, 2009.

\bibitem{sid13}
A.~Singh and J.~J. Dennehy, ``Stochastic holin expression can account for lysis
  time variation in the bacteriophage {\bf $\lambda$},'' \emph{Journal of the
  Royal Society Interface}, vol.~11, p. 20140140, 2014.

\bibitem{rao08}
A.~Raj and A.~van Oudenaarden, ``Nature, nurture, or chance: stochastic gene
  expression and its consequences,'' \emph{Cell}, vol. 135, pp. 216--226, 2008.

\bibitem{ulo16}
S.~Uphoff, N.~D. Lord, B.~Okumus, L.~Potvin-Trottier, D.~J. Sherratt, and
  J.~Paulsson, ``Stochastic activation of a dna damage response causes
  cell-to-cell mutation rate variation,'' \emph{Science}, vol. 351, pp.
  1094--1097, 2016.

\bibitem{nlp15}
T.~M. Norman, N.~D. Lord, J.~Paulsson, and R.~Losick, ``Stochastic switching of
  cell fate in microbes,'' \emph{Annual Review of Microbiology}, vol.~69, pp.
  381--403, 2015.

\bibitem{yxr06}
J.~Yu, J.~Xiao, X.~Ren, K.~Lao, and X.~S. Xie, ``Probing gene expression in
  live cells, one protein molecule at a time,'' \emph{Science}, vol. 311, pp.
  1600--1603, 2006.

\bibitem{zopf13}
C.~J. Zopf, K.~Quinn, J.~Zeidman, and N.~Maheshri, ``Cell-cycle dependence of
  transcription dominates noise in gene expression,'' \emph{PLOS Computational
  Biology}, vol.~9, p. e1003161, 2013.

\bibitem{sxn16}
S.~O. Skinner, H.~Xu, S.~Nagarkar-Jaiswal, P.~R. Freire, T.~P. Zwaka, and
  I.~Golding, ``Single-cell analysis of transcription kinetics across the cell
  cycle,'' \emph{eLife}, vol.~5, p. e12175, 2016.

\bibitem{pnb15}
O.~Padovan-Merhar, G.~P. Nair, A.~G. Biaesch, A.~Mayer, S.~Scarfone, S.~W.
  Foley, A.~R. Wu, L.~S. Churchman, A.~Singh, and A.~Raj, ``Single mammalian
  cells compensate for differences in cellular volume and {DNA} copy number
  through independent global transcriptional mechanisms,'' \emph{Molecular
  Cell}, vol.~58, pp. 339--352, 2015.

\bibitem{hup11}
D.~Huh and J.~Paulsson, ``Random partitioning of molecules at cell division,''
  \emph{Proceedings of the National Academy of Sciences}, vol. 108, pp.
  15\,004--15\,009, 2011.

\bibitem{lambert_2015}
G.~Lambert and E.~Kussell, ``Quantifying selective pressures driving bacterial
  evolution using lineage analysis,'' \emph{Physical Review X}, vol.~5, p.
  011016, 2015.

\bibitem{reshes_timing_2008}
G.~Reshes, S.~Vanounou, I.~Fishov, and M.~Feingold, ``Timing the start of
  division in e. coli: a single-cell study,'' \emph{Physical Biology}, vol.~5,
  p. 046001, 2008.

\bibitem{Roeder:2010vb}
A.~Roeder, V.~Chickarmane, B.~Obara, B.~Manjunath, and E.~M. Meyerowitz,
  ``{Variability in the control of cell division underlies sepal epidermal
  patterning in Arabidopsis thaliana},'' \emph{PLOS Biology}, vol.~8, p.
  e1000367, 2010.

\bibitem{Zilman:2010ud}
A.~Zilman, V.~Ganusov, and A.~Perelson, ``Stochastic models of lymphocyte
  proliferation and death,'' \emph{PlOS ONE}, vol.~5, p. e12775, 2010.

\bibitem{Hawkins:2009vw}
E.~D. Hawkins, J.~F. Markham, L.~P. McGuinness, and P.~Hodgkin, ``{A
  single-cell pedigree analysis of alternative stochastic lymphocyte fates},''
  \emph{Proceedings of the National Academy of Sciences}, vol. 106, pp.
  13\,457--13\,462, 2009.

\bibitem{sak13}
E.~B. Stukalin, I.~Aifuwa, J.~S. Kim, D.~Wirtz, and S.~Sun, ``{Age-dependent
  stochastic models for understanding population fluctuations in continuously
  cultured cells},'' \emph{Journal of the Royal Society Interface}, vol.~10, p.
  20130325, 2013.

\bibitem{gon13}
D.~Gonze, ``Modeling the effect of cell division on genetic oscillators,''
  \emph{Journal of Theoretical Biology}, vol. 325, pp. 22--33, 2013.

\bibitem{ltr14}
J.~Lloyd-Price, H.~Tran, and A.~S. Ribeiro, ``Dynamics of small genetic
  circuits subject to stochastic partitioning in cell division,'' \emph{Journal
  of Theoretical Biology}, vol. 356, pp. 11--19, 2014.

\bibitem{alo11}
U.~Alon, \emph{An Introduction to Systems Biology: Design Principles of
  Biological Circuits}.\hskip 1em plus 0.5em minus 0.4em\relax Chapman and
  Hall/CRC, 2011.

\bibitem{tcl10}
Y.~Taniguchi, P.~J. Choi, G.~W. Li, H.~Chen, M.~Babu, J.~Hearn, A.~Emili, and
  X.~S. Xie, ``Quantifying {E}. coli proteome and transcriptome with
  single-molecule sensitivity in single cells,'' \emph{Science}, vol. 329, pp.
  533--538, 2010.

\bibitem{sbl11}
B.~Schwanhausser, D.~Busse, N.~Li, G.~Dittmar, J.~Schuchhardt, J.~Wolf,
  W.~Chen, and M.~Selbach, ``Global quantification of mammalian gene expression
  control,'' \emph{Nature}, vol. 473, pp. 337--342, 2011.

\bibitem{gpz05}
I.~Golding, J.~Paulsson, S.~Zawilski, and E.~Cox, ``Real-time kinetics of gene
  activity in individual bacteria,'' \emph{Cell}, vol. 123, pp. 1025--1036,
  2005.

\bibitem{berg_1978}
O.~G. Berg, ``A model for the statistical fluctuations of protein numbers in a
  microbial population,'' \emph{Journal of Theoretical Biology}, vol.~71, pp.
  587--603, 1978.

\bibitem{rig79}
D.~R. Rigney, ``Stochastic model of constitutive protein levels in growing and
  dividing bacterial cells,'' \emph{Journal of Theoretical Biology}, vol.~76,
  pp. 453--480, 1979.

\bibitem{huh11}
D.~Huh and J.~Paulsson, ``Non-genetic heterogeneity from stochastic
  partitioning at cell division,'' \emph{Nature Genetics}, vol.~43, pp.
  95--100, 2011.

\bibitem{gov06}
C.~A. Gomez-Uribe and G.~C. Verghese, ``Mass fluctuation kinetics: Capturing
  stochastic effects in systems of chemical reactions through coupled
  mean-variance computations,'' \emph{Journal of Chemical Physics}, vol. 126,
  p. 024109, 2007.

\bibitem{lkk09}
C.~H. Lee, K.~Kim, and P.~Kim, ``A moment closure method for stochastic
  reaction networks,'' \emph{Journal of Chemical Physics}, vol. 130, p. 134107,
  2009.

\bibitem{gou07}
J.~Goutsias, ``Classical versus stochastic kinetics modeling of biochemical
  reaction systems,'' \emph{Biophysical Journal}, vol.~92, pp. 2350--2365,
  2007.

\bibitem{sih10}
A.~Singh and J.~P. Hespanha, ``Approximate moment dynamics for chemically
  reacting systems,'' \emph{IEEE Transactions on Automatic Control}, vol.~56,
  pp. 414--418, 2011.

\bibitem{gil09}
C.~S. Gillespie, ``Moment closure approximations for mass-action models,''
  \emph{IET Systems Biology}, vol.~3, pp. 52--58, 2009.

\bibitem{svs15}
M.~Soltani, C.~Vargas, and A.~Singh, ``Conditional moment closure schemes for
  studying stochastic dynamics of genetic circuits,'' \emph{IEEE Transactions
  on Biomedical Systems and Circuits}, vol.~9, pp. 518--526, 2015.

\bibitem{jdd14}
J.~Zhang, L.~DeVille, S.~Dhople, and A.~Dominguez-Garcia, ``A maximum entropy
  approach to the moment closure problem for stochastic hybrid systems at
  equilibrium,'' in \emph{IEEE Conference on Decision and Control}, 2014, pp.
  747--752.

\bibitem{sih10a}
A.~Singh and J.~P. Hespanha, ``Stochastic hybrid systems for studying
  biochemical processes,'' \emph{Philosophical Transactions of the Royal
  Society A}, vol. 368, pp. 4995--5011, 2010.

\bibitem{ses02}
P.~S. Swain, M.~B. Elowitz, and E.~D. Siggia, ``Intrinsic and extrinsic
  contributions to stochasticity in gene expression,'' \emph{Proceedings of the
  National Academy of Sciences}, vol.~99, pp. 12\,795--12\,800, 2002.

\bibitem{sis13}
A.~Singh and M.~Soltani, ``Quantifying intrinsic and extrinsic variability in
  stochastic gene expression models,'' \emph{PLOS ONE}, vol.~8, p. e84301,
  2013.

\bibitem{tha16}
M.~Thattai, ``Universal poisson statistics of mrnas with complex decay
  pathways,'' \emph{Biophysical Journal}, vol. 110, pp. 301--305, 2016.

\bibitem{boe13}
A.~N. Boettiger, ``Analytic {Approaches} to {Stochastic} {Gene} {Expression} in
  {Multicellular} {Systems},'' \emph{Biophysical Journal}, vol. 105, no.~12,
  pp. 2629--2640, 2013.

\bibitem{kmk09}
M.~Komorowski, J.~Miękisz, and A.~M. Kierzek, ``Translational repression
  contributes greater noise to gene expression than transcriptional
  repression,'' \emph{Biophysical Journal}, vol.~96, pp. 372--384, 2009.

\bibitem{rak08}
J.~Rausenberger and M.~Kollmann, ``Quantifying origins of cell-to-cell
  variations in gene expression,'' \emph{Biophysical Journal}, vol.~95, pp.
  4523--4528, 2008.

\bibitem{kis11}
K.~H. Kim and H.~M. Sauro, ``Measuring retroactivity from noise in gene
  regulatory networks,'' \emph{Biophysical Journal}, vol. 100, pp. 1167--1177,
  2011.

\bibitem{pau05}
J.~Paulsson, ``Model of stochastic gene expression,'' \emph{Physics of Life
  Reviews}, vol.~2, pp. 157--175, 2005.

\bibitem{sbf15}
M.~Soltani, P.~Bokes, Z.~Fox, and A.~Singh, ``Nonspecific transcription factor
  binding can reduce noise in the expression of downstream proteins,''
  \emph{Physical Biology}, vol.~12, p. 055002, 2015.

\bibitem{lll16}
Y.~Li, D.~Liu, C.~López-Paz, B.~J. Olson, and J.~G. Umen, ``A new class of
  cyclin dependent kinase in chlamydomonas is required for coupling cell size
  to cell division,'' \emph{eLife}, vol.~5, p. e10767, 2016.

\bibitem{stk15}
K.~M. Schmoller, J.~J. Turner, M.~Koivomagi, and J.~M. Skotheim, ``Dilution of
  the cell cycle inhibitor {Whi}5 controls budding-yeast cell size,''
  \emph{Nature}, vol. 526, pp. 268--272, 2015.

\bibitem{scs15}
K.~M. Schmoller and J.~M. Skotheim, ``The {Biosynthetic} {Basis} of {Cell}
  {Size} {Control},'' \emph{Trends in Cell Biology}, vol.~25, pp. 793--802,
  2015.

\bibitem{ama14}
A.~Amir, ``Cell size regulation in bacteria,'' \emph{Physical Review Letters},
  vol. 112, p. 208102, 2014.

\bibitem{gvs15}
K.~R. Ghusinga, C.~A. Vargas-Garcia, and A.~Singh, ``A mechanistic
  first-passage time framework for bacterial cell-division timing,''
  \emph{arXiv:1512.07864}, 2015.

\bibitem{vas16}
C.~A. Vargas-Garcia and A.~Singh, ``Hybrid systems approach to modeling
  stochastic dynamics of cell size,'' \emph{bioRxiv}, p. 044131, 2016.

\bibitem{mcq67}
D.~A. McQuarrie, ``Stochastic approach to chemical kinetics,'' \emph{Journal of
  Applied Probability}, vol.~4, pp. 413--478, 1967.

\bibitem{gil01}
D.~T. Gillespie, ``Approximate accelerated stochastic simulation of chemically
  reacting systems,'' \emph{Journal of Chemical Physics}, vol. 115, pp.
  1716--1733, 2001.

\bibitem{hsi04}
J.~P. Hespanha and A.~Singh, ``Stochastic models for chemically reacting
  systems using polynomial stochastic hybrid systems,'' \emph{International
  Journal of Robust and Nonlinear Control}, vol.~15, pp. 669--689, 2005.

\end{thebibliography}

\end{document}